# CROWDSOURCING COLLECTIVE EMOTIONAL INTELLIGENCE

Rob Morris & Rosalind Picard

Massachusetts Institute of Technology
77 Massachusetts Ave
Cambridge, Massachusetts, 02138, United States
rmorris@media.mit.edu, picard@media.mit.edu

**ABSTRACT**

One of the hallmarks of emotional intelligence is the ability to regulate emotions. While emotions can be regulated in many ways, a particularly powerful approach is cognitive reappraisal – a technique that involves reinterpreting the meaning of a thought or situation. Habitual use of this strategy is linked to many key indices of physical and emotional health, and laboratory studies show it can help down-regulate negative emotions, without incurring the physiological costs associated with other regulatory strategies. Unfortunately, cognitive reappraisal is not always easy to apply. Thinking flexibly about stressful thoughts and situations requires creativity and poise, faculties that often elude us when we need them the most. In this paper, we propose an assistive technology that coordinates collective intelligence on demand, to help individuals reappraise stressful thoughts and situations. In two experiments, we assess key features of our design and we demonstrate the feasibility of crowdsourcing empathetic reappraisals with on demand workforces, such as Amazon's Mechanical Turk.

**INTRODUCTION**

*"What really frightens or dismays us is not external events themselves, but the way in which we think about them."* – Epictetus

Over two thousand years ago, Epictetus, along with his Stoic contemporaries, anticipated one of the key insights of modern cognitive therapy: our thoughts play a crucial role in how situations affect us. By changing our cognitions, we can change our emotional responses. While there are many ways to achieve this, considerable research attention has been given to *cognitive reappraisal* - a technique that involves changing the meaning of a thought or situation to alter its emotional impact (Gross, 1998). To illustrate, consider a situation that is regrettably common in today's congested cities: a passing motorist cuts us off, honks loudly, and then gives us a one-fingered salute. Most of us would feel annoyed or outraged by this slight. Yet anger is by no means the inevitable outcome; we can always reappraise the situation to reduce our anger. For example, we might reappraise the driver's actions as ridiculously melodramatic, or even comical. Finding the humor in the situation could help take the edge off. Alternatively, we could think of the driver as someone deserving pity or compassion, not hate. As we dwell on this perspective, a sense of sadness might cut through our anger.

Some of these reappraisals may be more realistic or persuasive than others, but they are just a handful of the innumerable possibilities that exist. By thinking flexibly about stressful situations, we can alter our emotional experience in a number of different ways.

In recent years, cognitive reappraisal has been the topic of considerable psychological research. Neuroimaging research, for instance, has begun to outline the neural patterns associated with reappraisal (Goldin, Manber-Ball, Werner, Heimberg, & Gross, 2009; Ochsner & Gross, 2005). Research suggests that reappraisal can change both emotional experience and neurophysiological activation, affecting brain regions associated with emotional processing (e.g., amygdala and insula) and cognitive control (e.g., left prefrontal cortex and anterior cingulate cortex). Also, psychophysiological studies show that reappraisal can help regulate emotions, without incurring substantial physiological costs, such as increased skin conductance and heart rate (Gross, 1998; Jackson, Malmstadt, Larson, & Davidson, 2000). Further, psychological studies of individual differences link habitual reappraisal to healthy patterns of affect, interpersonal functioning, and subjective well-being (Gross & John, 2003). Researchers have also found a negative association between self-reported reappraisal use and rates of depression (Garnefski & Kraaij, 2006)

For many, however, cognitive reappraisal is not habitual, nor does it come naturally. While reappraisal can be taught by clinicians (and indeed, it is an important element in many psychotherapeutic traditions, such as CBT, DBT, and MBSR), not everyone has the time, money, or desire to pursue one-on-one counseling or psychotherapy. As such, assistive technologies that offer personalized reappraisal training and support could be extremely useful. For instance, an ideal new technology might be a mobile device that helps people adaptively reframe emotion-eliciting thoughts and situations. Such a tool could not only be a useful training aid, but it could also provide on-demand therapeutic support. For individuals with affective disorders, this tool could be a powerful adjunct to cognitive-behavioral therapy and/or dialectical-behavior therapy.

Building such a tool, however, would require enormous advances in artificial intelligence, natural language processing, affect analysis, computational psychology, and more. In this paper, we show how human-based computation, combined with the rise of online, crowdsourced workforces (e.g., Amazon's Mechanical Turk), can be used to create such a tool. What was previously an impossible computational task can now be accomplished with the aid of crowdsourced collective intelligence.

In the pages that follow, we describe ways to harness the collective emotional intelligence of online crowdworkers. Specifically, we outline a system that uses distributed human computation workflows to crowdsource emotion regulatory feedback. To our knowledge, our proposed system is the first to apply human computation techniques to the areas of emotional and mental health. Thus, our framework helps establish an important new point of intersection between the fields of collective intelligence and clinical and positive psychology.

In the pages that follow, we review related work in the fields of collective intelligence and we describe some contemporary, computer-based emotion support applications. Next, we outline a system that harnesses empathetic reappraisals using human computation approaches. Finally, in two experiments, we validate two important design elements and we demonstrate the feasibility of using a crowd-based system for: (1) empathizing, (2) detecting cognitive distortions, and (3) crafting relevant cognitive reappraisals.

## **RELATED WORK**

In our work, we incorporate techniques from two sub-disciplines within collective intelligence: crowdsourcing, and human computation. Given the definitional ambiguity of these fields (see Quinn & Bederson, 2011), we will clarify these terms for the purposes of this paper.

### **Crowdsourcing**

We define crowdsourcing as a method for recruiting and organizing collective intelligence. Following Howe's original definition of the term (Howe, 2006), we note that crowdsourcing usually involves an open-call for labor, without rigid contractual arrangements. For example, projects like Wikipedia, Linux, and TopCoder are all crowdsourced in the sense that anyone can contribute at any time, on a variety of different scales.

### **Human Computation**

Human computation, by contrast, relates more specifically to the type of work that gets done. Unlike large, peer production systems such as Wikipedia or Linux, human computation systems typically compress human labor into tight computational units which can then be organized and guided by machine-based systems and processes. For example, human computation games like *Fold-it*, *TagATune* and *ESP* coordinate human processing power in precise, circumscribed ways (Cooper et al., 2010; Law & von Ahn, 2009; L. von Ahn, 2006).

Together, crowdsourcing and human computation offer intriguing new ways to harness the collective intelligence of many different people. Systems that fall in this space come in many different forms and utilize many different motivational and computational structures to harness collective intelligence. For instance, human computation games use the intrinsically motivating properties of video games to recruit human volunteers (von Ahn, 2006). Other systems, like oDesk or Amazon's Mechanical Turk (MTurk), use monetary incentives to recruit workers. Workers may also be motivated to gain the respect and admiration of their peers. In general, most motivational strategies used in crowd-based, collective intelligence systems incentivize participation through some combination of love, glory, or money (see Malone, Laubacher, & Dellarocas, 2010).

Human computation systems can also differ in the way they coordinate human labor. Recently, new tools such as *TurKit*, *CrowdForge*, and *Jabberwocky* have given designers the ability to create increasingly complex human computation algorithms (Ahmad, Battle, Malkani, & Kamvar, 2011; Kittur, Smus, Khamkar, & Kraut, 2011; Little, Chilton, Goldman, & Miller, 2009). Instead of merely crowdsourcing tasks in parallel, these tools help designers build complex iterative workflows.

### **Crowd-Powered Interfaces**

Another new development in crowdsourced human computation is the emergence of crowd-powered, on demand interfaces. *Soylent* and *VizWiz*, for example, crowdsource human computation as needed, according to the actions of the end-user. In *Soylent*, a plugin for Microsoft Word, MTurk workers are recruited in a series of parallel and iterative stages to help users edit their documents (M. Bernstein et al., 2010). *VizWiz* uses similar methods to help blind users detect and locate items in their environment (Bigham et al., 2010). In *VizWiz*, a user uploads pictures of their surroundings, along with specific questions (e.g., "which can is the corn?"). The questions are handled immediately by MTurk workers, and responses are sent back to the user's phone. The system is a visual prosthetic, with multiple sets of crowdworkers providing eyes for the visually impaired end-users.

*VizWiz* and *Soylent* address specific challenges related to writing and visual perception, but the same methods could also be used to address a whole host of other cognition problems. In this paper, for example, we demonstrate that it is now feasible to crowdsource cognitive reappraisals using similar techniques.

### Online Emotional Support Tools

Anonymous, timely feedback of stressful life events is an important space, and several existing systems are working to fill this void. *Student Spill*, for example, uses a cadre of trained volunteers to provide empathetic and empowering responses to students. As of this writing, the service is only available at a handful of universities and responses can take up to 24 hours to be returned. *Emotional Bag Check*, by contrast, is open to anyone. Visitors to the site can choose to either vent their problems or address those of other users. The site primarily encourages responders to send mp3s to other users (*Emotional Bag Check* is self-described as "secretly a music site."), but support messages can also be sent. Both systems have their advantages, but there is room for improvement. To scale widely, systems cannot rely on a small cohort of trained volunteers, as in *Student Spill*. To provide therapeutic support, systems should guide responses according to evidence-based psychotherapeutic principles, unlike the open-ended framework of *Emotional Bag Check*.

In the next section, we describe a new system that could address many of these shortcomings. Specifically, we outline a new framework for crowdsourcing emotional support and cognitive reappraisal from an open pool of workers, using distributed human computation workflows.

To illustrate how this system might work, consider the following scenario: Michael, a 19-year-old college student, is starting a blog. He is excited about the project, but he finds it challenging, and he makes many mistakes. He opens an application on his phone and types the following: "I have been working on a blog and I have made many mistakes." The application asks him to describe the emotion(s) he feels and he notes that he is feeling "very stressed!" A few minutes later, he gets the following text from a real crowdworker: "*I'm sorry you are feeling stressed Michael. I understand how frustrating it can be when you just can't seem to get something right. Having to correct yourself like that can get tiring.*" Next, Michael receives a couple reappraisals. For instance, one says, "*Michael, anyone would feel stressed working on a blog, but not many people actually take the chance to write one.*" This short text reframes the situation from one of failure (making mistakes) to one of accomplishment (trying something challenging and daring). He continues to receive different reappraisals over the next few hours (quantity and delay can be limited).

This illustration reflects actual responses generated by our framework, in response to a real person's emotion-eliciting situation.[1] Our system is not yet fully automated end-to-end, and future work still needs to be done to build a robust user experience that can happen in real-time, but before we engineer real-time automation, we need to ensure that the crowdworkers can do this job properly. Care must be taken so that the support messages are well-composed and emotionally therapeutic. In the next section, we describe a workflow that we have tested that can now achieve this goal.

### DESIGN

Our system leverages MTurk to recruit workers and upload tasks (or "HITs," as they are often called on MTurk). Since our tasks require strong English language skills, we restrict enrollment to workers from the United States. We also limit our workforce to individuals with a 95% or higher approval rating. In pilot studies, we found these enrollment restrictions to dramatically increase the quality of our system's output.

### Workflow

Our system uses a distributed workflow, consisting of several parallel and iterative steps. The overarching task is broken into various subtasks, each of which gets sent to different groups of crowdworkers (see fig 1 for a visual overview of this framework). Disaggregating a complex task into smaller subtasks enables parallelization, which can reduce the overall latency of the system. Task instructions are also kept to a minimum, making it feasible to train workers on demand. In so doing, we eliminate the transaction

---

[1] While all our examples come from real people, we use fake names to preserve anonymity.

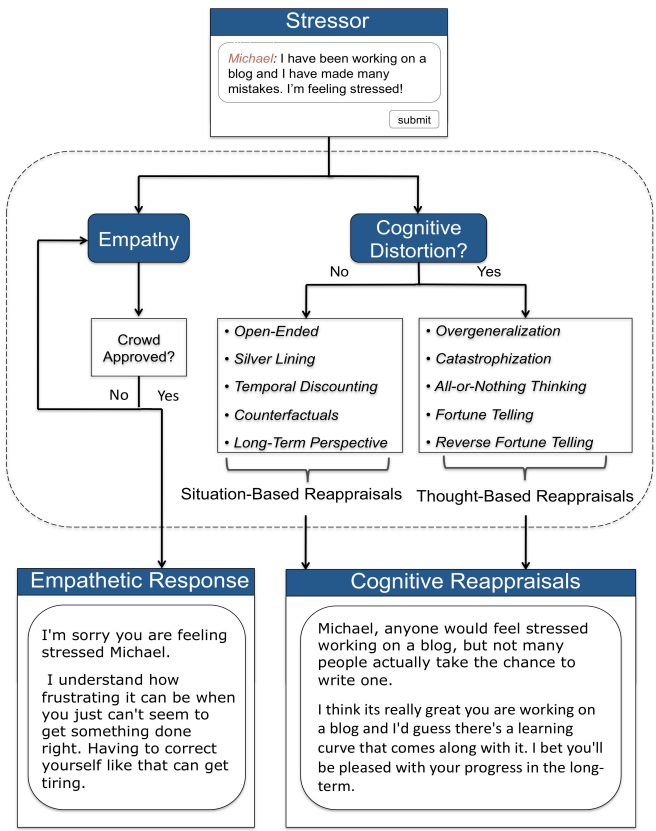

*Fig 1. Framework for crowdsourcing empathetic responses and cognitive reappraisals on MTurk. See text for more details.*

costs associated with retaining and managing a specialized pool of labor.[2]

**Input**

Users initiate the process by writing a one-to-three sentence description of a stressful thought and/or situation. By limiting the text entry to three sentences, we help users compartmentalize their stressors. Also, shorter text entries are easier to read and are therefore more manageable for the online workers.

**Empathy Task**

Once the user's text is sent to our system, two tasks are created in parallel – an empathy task and a classification task.

In the empathy task, a crowdworker sends the user a quick, empathetic response. The crowdworker is encouraged to use the following techniques in their response: (1) address the user directly, (e.g., "Michael, I'm sorry to hear …"), (2) let the user know that his/her emotion makes sense, given the situation, and (3) share how you might feel if you were in a similar situation. These techniques are inspired by research on emotional support messages (Greene & Burleson, 2003), and are designed to help workers craft effective empathetic responses.

We view the empathy response as a quick, first-line of socio-affective assistance. It helps comfort the user and it helps the user know that his/her concern is being addressed by real humans in our system.

In our research, we find that MTurk workers from the United States have little trouble generating empathetic responses, when they are instructed to do so (see experiment 1). Nonetheless, there is always the chance that a worker will misinterpret the instructions or provide unhelpful feedback. To address these concerns, other workers can be recruited to review the empathetic response before it gets sent to the user. If two workers agree that the response is appropriate, our system will pass it to the user. If not, another empathy HIT is created and a different worker is recruited to compose the response (see fig. 1 for a depiction of this feedback loop).

**Classification Task**

In parallel with the empathy task, different workers perform a binary classification on the input statement. This step helps guide our system towards contextually-relevant reappraisals. Here, workers are trained to determine whether or not the input statement includes cognitive distortions that might be addressable with thought-based reappraisal strategies. In cognitive therapy, cognitive distortions are defined as logical fallacies within negative statements (Beck, 1979). For example, the statement "*I'm going to flunk out of school and I'll never get a job, I know it!*" is distorted because it makes assumptions about the future that no one could know. There is no way this person could know that s/he will flunk out *and* be perpetually unemployed. By contrast, the statement "*There is construction on my street and I didn't get much sleep last night.*" is not distorted because it does not contain any illogical assumptions, predictions, or conclusions.

In the classification task, we quickly introduce our workers to the concept of cognitive distortions. We define the term and we show workers three examples of distorted statements and two examples of undistorted statements (see fig 2).

After this short training session, workers determine whether the user's text is or is not distorted. If a distortion is detected, the user's input statement is reframed using a technique called cognitive restructuring, as described below.

---

[2] For a longer discussion on task modularization in peer production systems and crowdsourcing markets, see (Benkler, 2002; Kittur et al., 2011).

Fig 2. Screenshot of the cognitive distortion tutorial.

**Thought-Based Reappraisals**

The same workers that detect cognitive distortions are also asked to reappraise them. Instead of passing the task onto another set of workers, we find it more efficient to retain the workers who have already completed the binary classification. These workers are already familiar with the user's input statement, and they are already trained on the concept of cognitive distortions. To reframe cognitive distortions, we simply ask these workers to gently explain why they think the user's statement might be distorted. Workers are given several labels of common cognitive distortions (see fig. 1), and they are asked to apply them if it seems appropriate to do so. This process is formally known as cognitive restructuring and is an important component in many schools of cognitive therapy (J. Beck, 2011).

While this reappraisal technique is often a bit foreign to some, we find that it is fairly easy to teach and apply. Workers are given some example responses for inspiration, but an extensive training session is not needed. To reframe cognitive distortions, workers simply need guidance on how to identify and repair distorted thinking.

**Situation-Based Reappraisals**

If no cognitive distortions are detected, crowdworkers attempt to reappraise the meaning or subjective interpretation of the user's situation. Workers are specifically told not to give advice or suggest ways to solve the problem. Rather, they are instructed to suggest different ways of thinking about the user's situation. After a quick introduction to the concept, workers are given examples of good and bad reappraisals. The examples are designed to dissuade workers from making two common mistakes we frequently observe: offering advice and making unrealistic assumptions about the user's situation.

We also ask workers to limit their responses to four sentences. This helps eliminate problems caused by well-intentioned but over-zealous workers (see Bernstein et al, 2010 for a description of the "eager beaver" MTurk worker).

In our system, some workers are simply asked to come up with their own reappraisal suggestions. These workers are not told to use any particular reappraisal techniques in their responses. Using this approach, we often see a modal reappraisal emerge. That is, many workers will independently come up with the same way to reinterpret a situation. For the user, this redundancy could add legitimacy to the reappraisal - a crowd of strangers has agreed on a more beneficial way to view a stressful situation. However, some users might also like variety, to help them reconsider situations in new and unexpected ways.

To encourage this variety, therefore, another set of workers is asked to try specific reappraisal strategies that might be less obvious. For example, some workers are asked to find potential silver linings in the situation, while others are asked to assess the situation from a long-term perspective (see fig. 1 for a list of some of these reappraisal strategies). These instructions help guide workers towards reappraisals that might not initially be considered by the other crowdworkers.

Before we deploy an automated system in real user studies, we first need to test our primary design assumptions. In the pages that follow, we describe two experiments that evaluate two key elements of our system.

**EXPERIMENT 1: RESPONSE QUALITY**

Our system assumes that high quality cognitive reappraisals and empathetic statements will not arise naturally from the crowd. However, perhaps this is an overly pessimistic view of crowdworkers. If most crowdworkers naturally generate good responses, then human computation algorithms would not need to guide workers or control for quality. To explore this possibility, we compared responses from two separate conditions: an *unstructured* condition, in which workers were simply asked to help the user feel better and a *structured* condition, in which workers were asked to provide empathetic statements and cognitive reappraisals. The latter condition utilized several of the crowdflow algorithms described in the previous section.

**Method**

In our experiment, participants were asked to respond to three input statements, previously supplied by other MTurk workers (see Table 1).

After accepting our HIT, 102 participants were randomly assigned to the unstructured or structured condition. In the unstructured condition, participants were simply asked to help the target feel better about his/her situation. They were asked to limit their response to six sentences or less. In the structured condition, participants were asked to first empathize with the target, using no more than three sentences. They were then asked to help the target reframe the situation to make it seem less distressing. For the reappraisal component, responses were also limited to three sentences. As such, the total length of the structured and unstructured responses was balanced and limited to six sentences in both conditions.

| |
|---|
| Michael says, *"I have been working on a blog and have made many mistakes. I'm feeling really stressed."* |
| Sarah says, *"My boyfriend did not call me this morning, like he said he would. I'm feeling really angry"* |
| Jack says, *"Yesterday my dad drank the last of the coffee and didn't make more. I'm feeling really irritated!"* |

*Table 1. The three input statements used in experiment 1.*

Next, we recruited 70 MTurk workers to rate the responses. Our raters saw a random mixture of 34 structured and unstructured responses. We also included four decoy responses, two of which were off-topic, and two of which were overtly rude and uncaring. Five raters failed to respond appropriately to the decoy responses and were not included in the overall ratings scores.

For each response, workers were asked to rate the extent to which they agreed or disagreed with the following two statements:

1) *This response is empathetic. The responder seems to sympathize with this individual's situation.*

2) *This response offers a positive way to think about this situation.*

Ratings were made using a 7-point likert scale, with endpoints labeled as 1="strongly disagree" and 7="strongly agree." We used data from the first and second likert questions as scores for empathy and reappraisal, respectively.

### Results

To examine the difference between the structured and unstructured responses, we ran a two-way MANOVA, with response structure (structured vs. unstructured) as a between-subjects factor. Empathy and reappraisal scores were used as our dependent variables, and we set the type of input stressor as a covariate in our analyses.

In support of our hypothesis, we found that empathy scores were significantly higher in the structured condition ($M = 5.71$, $SD = .62$) compared to the unstructured condition ($M = 4.14$, $SD = 1.21$), [$F(1, 99) = 73.02$, $p < .005$]. Similarly, the structured condition had significantly higher reappraisal scores ($M = 5.45$, $SD = .59$) than the unstructured condition ($M = 4.41$, $SD = 1.11$), [$F(1, 99) = 34.90$, $p < .005$] Our covariate analysis showed no significant effect of input statement on either the empathy scores [$F(1, 99) = .387$, $p > .54$] or reappraisal scores [$F(1, 99) = .194$, $p > .66$], suggesting that the type of stressful situation did not produce differential effects across the different conditions.

### Discussion

Our results support the hypothesis that, with guidance, crowdworkers will help craft empathetic reappraisals for strangers. By contrast, when told to simply make a person feel better, crowdworkers are less likely to be empathetic and offer reappraisals.

In both the structured and unstructured conditions, participants were told their responses would be sent to real people. We assumed this would prompt the vast majority of workers to respond with an empathetic statement. Yet, responses in the unstructured condition were considerably less empathetic than responses in the structured condition. While research shows that emotional support messages should usually include empathetic components (Greene & Burleson, 2003), workers did not naturally include these in their responses. Further, responses from the unstructured condition were less likely to include convincing reappraisals. In short, the crowdsourced workers' responses were more emotionally intelligent when they were guided by our system.

### EXPERIMENT 2: CLASSIFICATION

Our design also assumes that crowdworkers can reliably identify cognitive distortions, with very little training. In our second experiment, we test this assumption by asking workers to classify a set of input statements as either distorted or undistorted.

### Method

We recruited 73 participants from Amazon's Mechanical Turk. Participants were briefly trained to detect cognitive distortions, as described previously, and they were asked to classify a set of 32 input statements.

Each statement was negatively valenced and included a one-to-three sentence description of an emotion-eliciting thought or situation. Half of the statements were distorted in some way. The cognitive distortions were taken from online resources[3] and cognitive therapy literature (Burns, 1999), and were chosen to capture a wide variety of distorted thoughts (see Table 2). The undistorted, statements included descriptions of real stressors submitted by MTurk workers. Each participant saw a random mixture of distorted and undistorted statements. Participants were told to read each statement and classify it as distorted or undistorted.

---

[3] http://www.drbeckham.com/handouts/

| Classification | Input Statements |
|---|---|
| Distorted | *"My son acted up at church. Everyone must think I have no control over him and that I'm a terrible parent."* <br> *"I forgot my lines in the play and really made a fool of myself."* |
| Undistorted | *"My best friend doesn't call me as much as she used to."* <br> *"My car needs to be repaired, but I'd rather use that money to pay my rent!"* |

*Table 2. Examples of the distorted and undistorted statements workers were asked to classify.*

### Results

We created a confusion matrix to plot MTurk classifications against the ground truth. We calculated accuracy by dividing the number of correct classifications (true positives and true negatives) by the total number of all recorded classifications. On average, MTurk workers correctly classified 89% (*SD*=7%) of the input statements (see fig. 3).

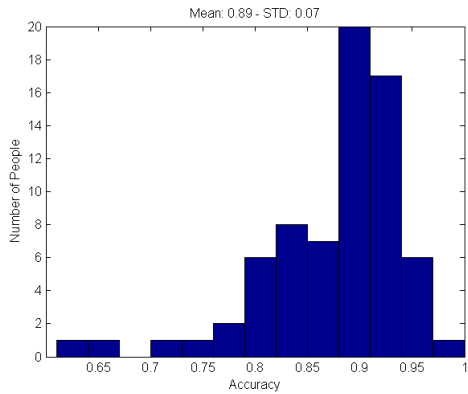

*Fig 3. A histogram of classification accuracy from our sample of MTurk workers.*

### Discussion

Based on our results from experiment 2, we conclude that MTurk workers can reliably identify cognitive distortions within short, one-to-three sentence descriptions. With minimal instructions, MTurk workers seemed to understand the concept. Future work might explore how to leverage this classification procedure for other applications. For instance, the number of cognitive distortions detected in online communications or phone transcripts might be an important diagnostic indicator of affective disorders.

Taken together, the results from experiments 1 and 2 suggest that we have the main design components required to put together a crowd-based system for: (1) empathizing, (2) detecting cognitive distortions, and (3) crafting cognitive reappraisals.

### FUTURE WORK

Future work involves automating our system end-to-end and deploying it with a real user population. Additional steps will be required to reduce the latency of the system. Our current design does not yet incorporate state-of-the-art, rapid crowdsourcing techniques, such as automatic task re-posting and worker retainer systems (Bernstein, Brandt, Miller, & Karger, 2011; Bigham et al., 2010). Applying these techniques should help the system respond faster, without sacrificing quality.

To improve the overall user experience, users should be able to rate the helpfulness of the responses they receive. Over time, the system could start to learn which types of reappraisals work best for different users and for different categories of problems. This could help the system apply person-specific and situation-specific response algorithms. Ideally, the tone of the responses should also be tailored to the personality profile of the user. For example, research by Nass and colleagues illustrates how human-computer interactions can be improved by matching the technology's personality with the personality of the end-user (Nass et al., 1995). Future versions of our system might guide workers to write more or less submissive or assertive reappraisals, depending on the needs and personality of the user.

Future work also involves researching potential long-term therapeutic effects of this system. We are interested in whether crowd-based feedback has any unique therapeutic properties. For example, in traditional cognitive therapeutic settings, therapists often teach patients to question negative thought patterns through a process known as "collaborative empiricism." A crowd-based system, by contrast, might involve "collective empiricism," – an approach where crowdworkers, not therapists, question the veracity of a user's thoughts and appraisals and offer new perspectives. We believe that crowd-based feedback could have unique persuasive power, and we hope to explore this in future experiments.

Finally, we would like to explore the effects of being a contributor in the kind of system we envision. It may be very useful for individuals to practice reappraising the thoughts and situations of other people. If this is the case, then it might behoove us to move beyond micro-task markets, where incentives are largely pecuniary, and instead consider a system built on reciprocity, where users are also contributors.

### CONCLUSION

This paper presents a new way to crowdsource cognitive reappraisal – a key emotion regulatory strategy that is linked to emotional health and emotional intelligence. We propose a system that uses a combination of empathy, cognitive therapeutic techniques, and the combined insights of many different people. Our experiments demonstrate the feasibility of this system and suggest that human computation algorithms can improve the quality of crowdsourced empathetic reappraisals. Future work involves fully automating and optimizing the system for different kinds of inputs. We also believe that this new kind of system, and variants thereof, could stimulate new research on the ramifications of crowd-based cognitive therapeutic interventions.

**ACKNOWLEDGEMENTS**

We would like to thank Mira Dontcheva, Laura Ramsey, Javier Hernandez, and the Stanford Clinically Applied Affective Neuroscience group for valuable feedback on this work. This work was funded in part by the MIT Media Lab Consortium.